%% file: main.tex
\documentclass[sigconf]{acmart}
\usepackage{algorithmicx}
\usepackage{algorithm}
\usepackage{amsmath}
\usepackage{enumitem}
\usepackage{algpseudocode}
\usepackage{multirow}
\usepackage{epsfig}
\usepackage{multicol,lipsum,xparse}
\usepackage{subcaption}
\usepackage{amsmath}
\usepackage{stmaryrd}
\usepackage{color}
\usepackage{todonotes}
\usepackage[utf8]{inputenc}

\usepackage{tikz}

\title{Developing Synthesis Flows Without Human Knowledge}
\vspace{4mm}

\author{Cunxi Yu}
\affiliation{%
  \institution{Integrated Systems Laboratory, EPFL}
  \city{Lausanne}
  \state{Switzerland}
}
\email{cunxi.yu@epfl.ch}

\author{Houping Xiao}
\affiliation{%
  \institution{SUNY Buffalo}
  \city{Buffalo, NY}
  \state{USA}
}
\email{houpingx@buffalo.edu }

\author{Giovanni De Micheli}
\affiliation{%
  \institution{Integrated Systems Laboratory, EPFL}
  \city{Lausanne}
  \state{Switzerland}
}
\email{giovanni.demicheli@epfl.ch}

\begin{document}

\settopmatter{printacmref=false}

\newcommand{\pgftextcircled}[1]{
    \setbox0=\hbox{#1}%
    \dimen0\wd0%
    \divide\dimen0 by 2%
    \begin{tikzpicture}[baseline=(a.base)]%
        \useasboundingbox (-\the\dimen0,0pt) rectangle (\the\dimen0,1pt);
        \node[circle,draw,outer sep=0pt,inner sep=0.1ex] (a) {#1};
    \end{tikzpicture}
}

\copyrightyear{2018} 
\acmYear{2018} 
\setcopyright{acmcopyright}
\acmConference[DAC '18]{DAC '18: The 55th Annual Design Automation Conference 2018}{June 24--29, 2018}{San Francisco, CA, USA}
\acmBooktitle{DAC '18: DAC '18: The 55th Annual Design Automation Conference 2018, June 24--29, 2018, San Francisco, CA, USA}
\acmPrice{15.00}
\acmDOI{10.1145/3195970.3196026}
\acmISBN{978-1-4503-5700-5/18/06}

\renewcommand{\shortauthors}{C. Yu et al.}

\begin{abstract}

Design flows are the explicit combinations of design transformations, primarily involved in synthesis, placement and routing processes, to accomplish the design of Integrated Circuits (ICs) and System-on-Chip (SoC). Mostly, the flows are developed based on the knowledge of the experts. However, due to the large search space of design flows and the increasing design complexity, developing \textit{Intellectual Property (IP)-specific} synthesis flows providing high Quality of Result (QoR) is extremely challenging. This work presents a fully autonomous framework that artificially produces design-specific synthesis flows without human guidance and baseline flows, using Convolutional Neural Network (CNN). The demonstrations are made by successfully designing logic synthesis flows of three large scaled designs.

\end{abstract}

\maketitle

\input{intro.tex}
\input{background.tex}
\input{approach.tex}

\input{results.tex}

\scriptsize
\bibliographystyle{IEEEtran}
\bibliography{synthesis}

\end{document}

%% file: intro.tex
\section{Introduction}

Electronic Design Automation (EDA) involves a diverse set of software algorithms and applications that are required for the design of complex electronic systems. Given the deep design challenges that the designers are facing, developing high-quality and efficient design flows has been crucial. A well-developed design flow could reduce time-to-market by enabling manufacturability, addressing timing closure and power consumption, etc. In general, the EDA vendors provide reference design flows along with the EDA tools. However, such design flows may not perform well for many designs.

\begin{figure*}[!htb]
\centering
\begin{minipage}{0.235\textwidth}
  \centering
  \small
\includegraphics[width=1\textwidth]{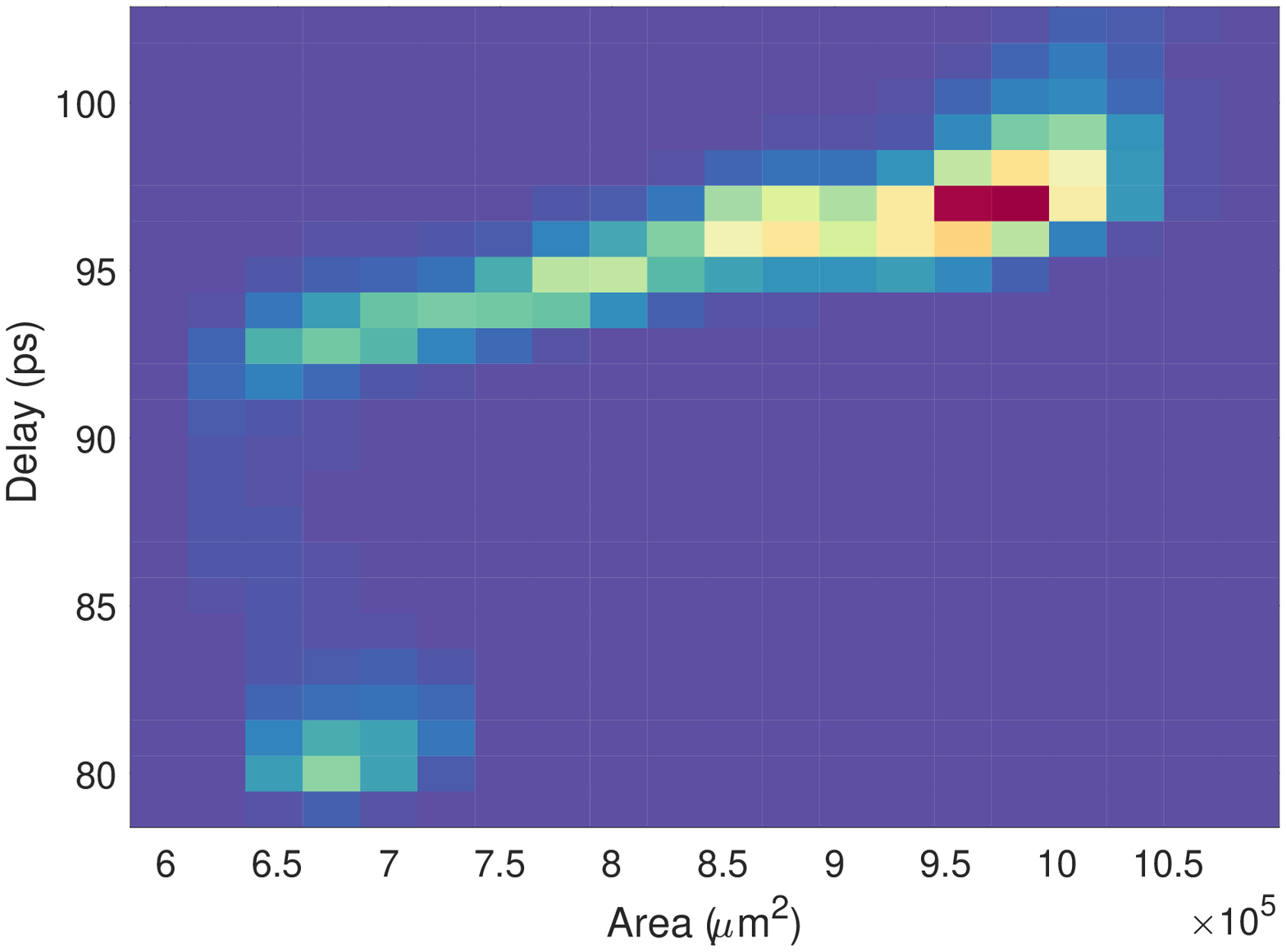}
\subcaption{2-D QoR distro of AES}\label{fig:motivation1-aes}
\end{minipage}%
\begin{minipage}{0.235\textwidth}
  \centering
\includegraphics[width=1\textwidth]{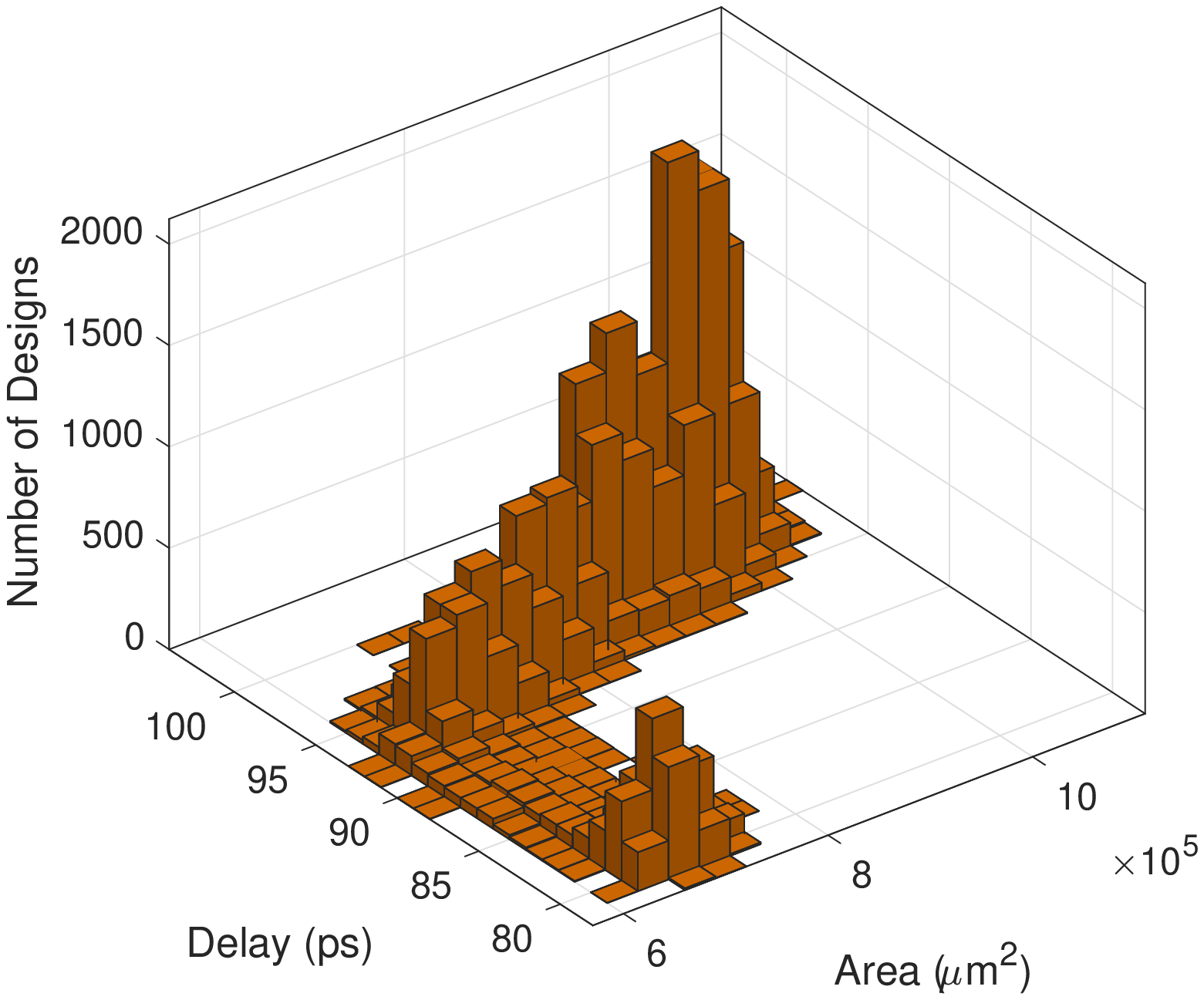}
\subcaption{3-D QoR distro of AES}\label{fig:motivation2-aes}
\end{minipage}%
\begin{minipage}{0.235\textwidth}
  \centering
\includegraphics[width=1\textwidth]{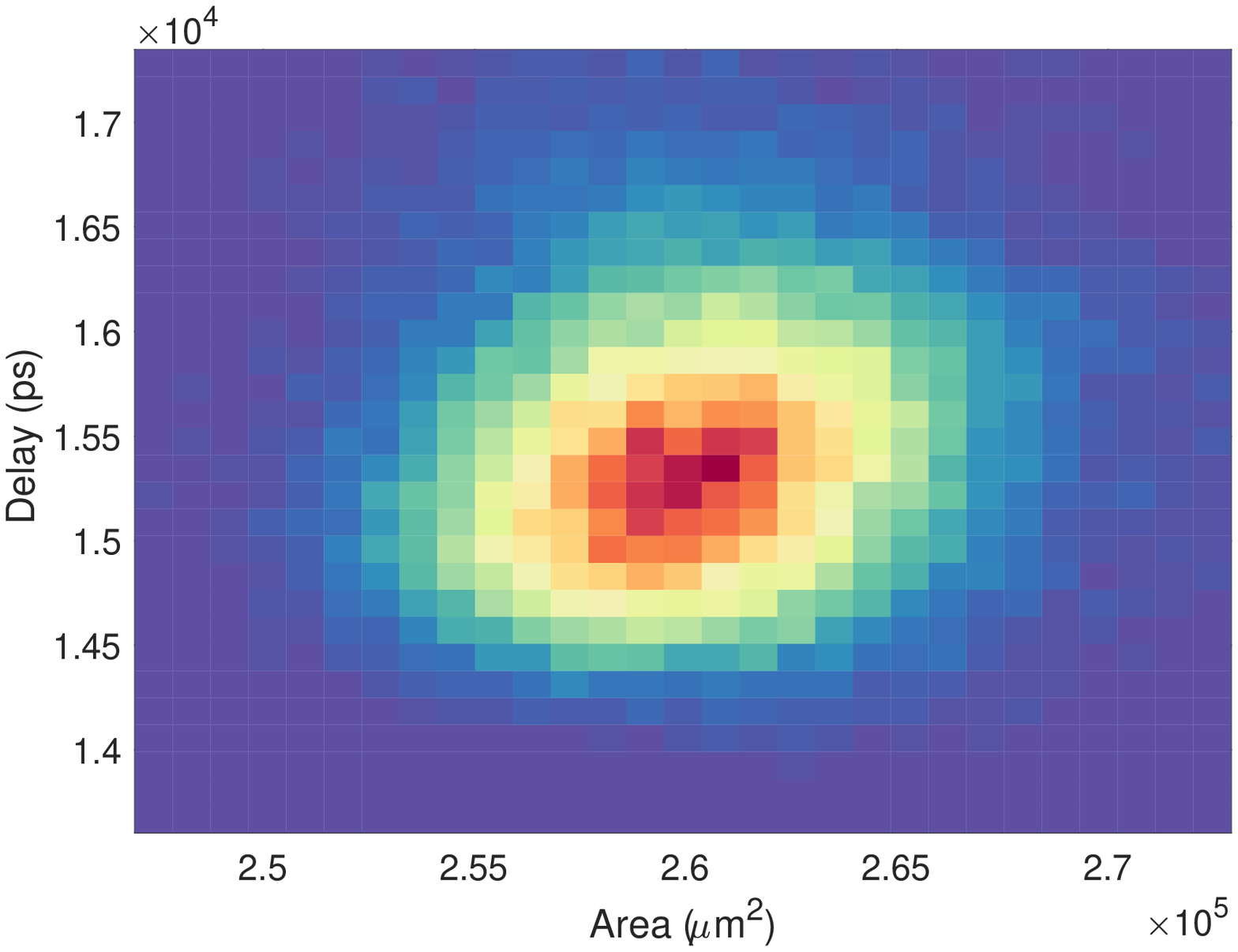}
\subcaption{2-D QoR distro of ALU}\label{fig:motivation1-datapath}
\end{minipage}%
\begin{minipage}{0.235\textwidth}
  \centering
\includegraphics[width=1\textwidth]{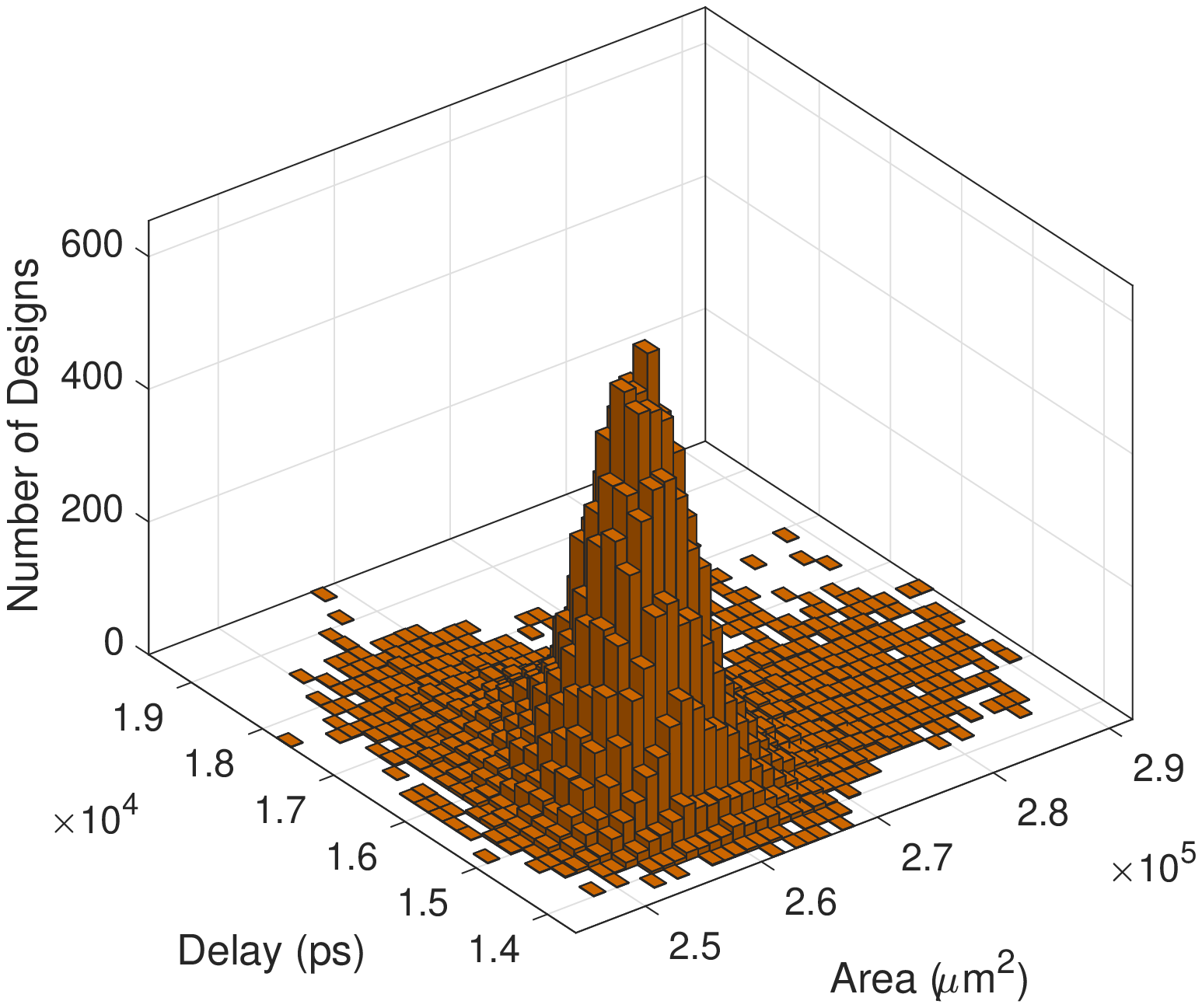}
\subcaption{3-D QoR distro of ALU}\label{fig:motivation2-datapath}
\end{minipage}%
\caption{Delay and area results of the 50,000 random ABC synthesis flows of 128-bit AES core and 64-bit ALU.}
\vspace{-4mm}
\label{fig:motivation}
\end{figure*}

There are two major reasons. First, the performance of the design flow varies on the Intellectual Property (IP) of the design. To achieve the design objectives, design flows need to be customized for the given IP. Such flows are called \textit{IP-specific} or \textit{design-specific} flows. This becomes more important while new types of designs are coming out, e.g., design methods for Neuromorphic chip \cite{akopyan2015truenorth}. Second, the design flows are mostly developed by the EDA developers and users based on their knowledge and user experience, with many testing iterations and intensive supervision. However, due to a large number of available flows, finding the best design flows among the entire search space by human-testing is impossible. It is particularly difficult to find the best flows for the recently developed transformations \cite{cunxiyu:dac16,YuHNCKSCM18}. For example, given 50 synthesis transformation that each of them can be processed independently. The total number of available design flows is $50!$ $\approx$ $3 \cdot 10^{64}$. The search space of general flows is formally defined in Section 2.1. Although the significant efforts spent in providing high-quality design flows, the technique that systematically generates \textit{IP-specific} synthesis flows has been lagging. Similarly, these problems exist in designing System-on-Chip (SoC). In Section 2 (Figure \ref{fig:motivation}), two motivating examples are provided to show the needs of developing such technique.

Design flows are considered as iterative flows since the transformations are applied to the designs iteratively. \textit{Machine learning} technique has been leveraged in flow optimization, such as iterative flow optimization for compilers using \textit{Markov Chain} \cite{agakov2006using}. Regarding synthesis flow optimization, Liu et al. recently introduced an area optimization approach for Look-up-table (LUT) mapping, in which the logic transformations are guided using \textit{Markov Chain Monte Carlo} (MCMC) method \cite{liu2017parallelized}. However, Markov Chain model is not sufficient in autonomously designing synthesis flows. The main reason is that the synthesis transformation(s) may not affect the next transformation but affect the transformation several iterations later, which does not satisfy the \textit{Markov Property} \cite{durrett2010probability}. In this work, we formulate the problems of artificially developing synthesis flows as a \textit{Multiclass classification} problem, and solved using \textit{Deep learning} \cite{lecun2015deep}. \textit{Deep learning} has shown considerable success in tasks like image recognition \cite{krizhevsky2012imagenet} and natural language processing \cite{farabet2013learning}. Several advances mitigate the deficiencies of traditional multilayer perceptrons (MLPs), e.g., CNNs have made it possible to robustly and automatically extract learned features; \textit{over-fitting} is mitigated in fully connected layers using the random regularization called dropout \cite{srivastava2014dropout}.

Specifically, this paper includes the following contributions: \textbf{a)} The search space of artificially developing synthesis flows is formally defined in Section 2. \textbf{b)} We introduce a flow-classification model (Section 3.1) combining with the one-hot modeling of flows (Section 3.2), such that the problem can be modeled as \textit{Multiclass Classfication} problem. \textbf{c)} We develop a fully autonomous framework for developing synthesis flows based on Convolutional Neural Network (CNN). This framework takes HDL as input and output two sets of synthesis flows, namely \textit{angel}-flows and \textit{devil}-flows that provide the best and worst QoRs respectively\footnote{\small \textit{devil}-flows could provide information for improving the synthesis transformations.}. \textbf{d)} Our framework is demonstrated by successively developing delay-driven and area-driven \textit{angel/devil}-flows for 64-bit Montgomery Multiplier, 128-bit AES core and 64-bit ALU. Evaluations of the CNN architecture and training process for classifying synthesis flows are also provided. \textbf{e)} The datasets and demos are released publicly\footnote{https://github.com/ycunxi/FLowGen-CNNs-DAC18.git}. 

%% file: background.tex

\section{Background}

\subsection{Notations and Search Space}

\noindent
\textbf{Definition 1 \textit{none}-repetition Synthesis Flow}: \textit{Given a set of unique synthesis transformations $\mathbb{S}$=\{$p_0$, $p_1$,..., $p_n$\}, a synthesis flow $\mathbb{F}$ is a permutation of $p_i$ $\in$ $\mathbb{S}$ performed iteratively.}

\textbf{Example 1:} Let $\mathbb{S}$=\{$p_0$, $p_1$, $p_2$\}. $p_i$ are the transformations in the synthesis tools and can be processed independently. Then, there are totally six flows available: 

\vspace{0mm}
\begin{equation*}
\scriptsize
\begin{aligned}
F_0 : p_0 \rightarrow p_1 \rightarrow p_2 ~~
F_1 : p_0 \rightarrow p_2 \rightarrow p_1 \\
F_2 : p_1 \rightarrow p_0 \rightarrow p_2 ~~
F_3 : p_1 \rightarrow p_2 \rightarrow p_0 \\
F_4 : p_2 \rightarrow p_0 \rightarrow p_1 ~~
F_5 : p_2 \rightarrow p_1 \rightarrow p_0
\end{aligned}
\end{equation*}

\noindent
\textbf{Remark 1}: \textit{Let $\mathbb{N}$ be the number of all available flows, where $\mathbb{S}$ includes $n$ elements, such that $\mathbb{N}$ $\leq$ $n!$. }

The upper bound of $\mathbb{N}$ happens \textit{iff} all elements in $\mathbb{S}$ can be processed independently. In practice, there could be some constraints have to be satisfied for processing these transformations. In this case, $\mathbb{N}$ will be smaller than $n!$. For example, given a constraint that $p_1$ has to be processed before $p_2$, the available flows include only $F_0$, $F_2$, and $F_3$.

\noindent
\textbf{Definition 2 $m$-repetition Synthesis Flow (m$\geq$2)}: \textit{Given a set of unique synthesis transformations $\mathbb{S}$=\{$p_0$, $p_1$,...,$p_n$\}, a synthesis flow with $m$-repetition $\mathbb{F}_{m}$ is a permutation of $p_i$ $\in$ $\mathbb{S}_{m}$, where $\mathbb{S}_{m}$ contains $m$ $\mathbb{S}$ sets.} 

\textbf{Example 2:} Let $\mathbb{S}$=\{$p_0$, $p_1$\}. Each $p_i$ can be processed independently. For developing $2$-repetition synthesis flows, $\mathbb{S}_{2}$=\{$p_0$, $p_1$, $p_0$, $p_1$\}. The available flows include:

\vspace{0mm}
\begin{equation*}
\scriptsize
\begin{aligned}
F_0 : p_0 \rightarrow p_0 \rightarrow p_1 \rightarrow p_1~~
F_1 : p_1 \rightarrow p_1 \rightarrow p_0 \rightarrow p_0 \\
F_2 : p_0 \rightarrow p_1 \rightarrow p_0 \rightarrow p_1~~
F_3 : p_1 \rightarrow p_0 \rightarrow p_1 \rightarrow p_0 \\
F_4 : p_0 \rightarrow p_1 \rightarrow p_1 \rightarrow p_0~~
F_5 : p_1 \rightarrow p_0 \rightarrow p_0 \rightarrow p_1 \\
\end{aligned}
\end{equation*}

\noindent
\textbf{Remark 2}: \textit{Let $\mathbb{L}$ be the length of a synthesis flow. Given a  $m$-repetition $\mathbb{F}_{m}$ with $n$ transformations in $\mathbb{S}$, $\mathbb{L}$ = $n$$\times$$m$.}

\noindent
\textbf{Remark 3}: \textit{Let function $f(n,\mathbb{L},m)$ be the number of available $m$-repetition flows with $n$ elements in $S$. $f(n,\mathbb{L},m)$ uniquely satisfies the following recursive formula :}

\vspace{0mm}
\begin{equation*}
\scriptsize
\begin{aligned}
f(n,\mathbb{L}+1,m) = n f(n,\mathbb{L},m) - n \binom{\mathbb{L}}{m} f(n-1,\mathbb{L}-m,m)\\
\end{aligned}
\end{equation*}

{The number of available $m$-repetition flows with $n$ synthesis transformations is the same as counting $\mathbb{L}$-\textit{permutations} of $n$ objects. The proof of the recursive formula is similar to \cite{mendelson1981permutations} that will not be included in this paper. The upper and lower boundary conditions are $n!$ < $f(n,\mathbb{L},m)$ < $n^{\mathbb{L}}$. We can see that $f(n,\mathbb{L},m)$ becomes dramatically larger than $n!$ (\textit{non}-repetition flows) as $m$ increasing.}

\subsection{Motivating Example}\label{sec:motivating-example}

We provide two motivating examples using the Open-source logic synthesis framework ABC \cite{mishchenko2010abc} shown in Figure \ref{fig:motivation}. The setups are as follows:
    \vspace{0mm}
\begin{itemize}[leftmargin=*]
    \item $\mathbb{S}$=\{balance, restructure, rewrite, refactor, rewrite -z, refactor -z\} ($n$=6); the elements in  $\mathbb{S}$ are logic transformations in ABC\footnote{\small The names of these transformations are the same as the commands in ABC.} that can be processed independently. 
        \vspace{0mm}
    \item 50,000 unique $4$-repetition flows are generated by random permutations of  $\mathbb{S}_{4}$ ($m$=4, $n$=6, $\mathbb{L}$=24).
        \vspace{0mm}
    \item Input designs: 128-bit Advanced Encryption Standard (AES) core, and 64-bit ALU taken from OpenCore \cite{opencore-web}.
        \vspace{0mm}
    \item Delay and area of these flows are obtained after technology mapping using a 14nm standard-cell library.
        \vspace{0mm}
\end{itemize}

The QoR distributions of AES and ALU designs using the 50,000 random flows are shown in Figure \ref{fig:motivation}-(a, b) and (c, d). There are several important observations based on Figure 1, which show the main motivations of this work:
    \vspace{0mm}

\begin{itemize}[leftmargin=*]
    \item Given the same set of synthesis transformations, the QoR is very different using different flows. For example, delay and area of AES design produced by the 50,000 flows have up to 40\% and 90\% difference, respectively. 
    \vspace{0mm}
    \item The search space of the synthesis flows is large. According to Remark 3, the total number of available $4$-repetition flows with $n=6$ independent synthesis transformations is more than $10^{16}$. Discovering the high-quality synthesis flows with human-testing among the entire search space is unlikely to be achieved.
        \vspace{0mm}
    \item The same set of flows perform differently on different designs. For example, in Figure \ref{fig:motivation}, QoR distributions of AES and ALU are \textit{statistically significant}. This means that the high-quality flows for AES design could perform poorly for ALU. Therefore, synthesis flows need to be customized for specific IP or design.
        \vspace{0mm}
    
\end{itemize}

%% file: approach.tex
\section{Approach}

\begin{figure}[!htb]
\centering
\includegraphics[width=.43\textwidth]{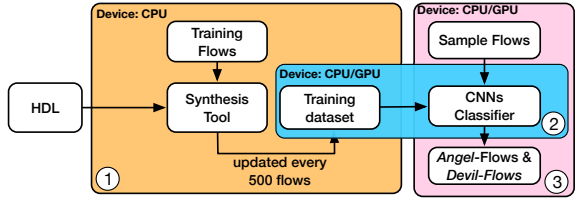}
\caption{Overview of the proposed framework, performing in sequence \textcircled{1} $\rightarrow$ \textcircled{2} $\rightarrow$ \textcircled{3}.}
\vspace{-1mm}
\label{fig:flow}
\end{figure}

\subsection{Overview}\label{sec:overview}

This section presents our framework that artificially develops synthesis flows for a given design. Our framework takes the HDL as input and outputs two sets of synthesis flows, namely \textit{angel-flows} and \textit{devil-flows}, which provide the best and worst QoR according to the design objectives. This problem is formulated as \textit{Multiclass Classification} and solved using CNN classifier. The main idea of our approach is that training a CNN Classifier with a small set of \textit{labeled} random flows. The classes (or labels) of the synthesis flows are labeled based on one or multiple QoR metrics, such as delay, area, power, etc. The trained classifier is used to predict the classes of a large number of \textit{unlabeled} random flows. Finally, \textit{angel-flows} and \textit{devil-flows} are generated by sorting the \textit{prediction confidence}, i.e., the probability to be in a certain class (Section \ref{sec:angel-devil}). This framework is a generic model for designing synthesis flows in many stages, such as High-level synthesis and logic synthesis. The demonstration is made by designing logic synthesis flows using ABC \cite{mishchenko2010abc} shown in Section \ref{sec:results}. The flow of our framework is shown in Figure \ref{fig:flow}, including three main components:

\textbf{\textcircled{1} Generate training datasets.} In this work, the training dataset is a set of \textit{labeled} synthesis flows, namely \textit{training flows}. However, the training flows are originally \textit{unlabeled}. This first step of our approach is labeling a set of random flows. This requires applying these synthesis flows to the input design and collecting the QoR result at the end of each flow. 
Note that applying a synthesis flow to a large design could be time-consuming. Hence, our framework is performed in an incremental fashion. The CNN training (component \textcircled{2}) starts after 1000 labeled flows collected, and it will be re-trained every 500 new labeled flows collected. In this case, our framework can produce the intermediate results during the training process.

These flows will be labeled according to the classification model shown in Table \ref{tbl:labeling}. This model can be changed according to the design objectives, using either a single-metric or multi-metric model. For example, if the design objective is area optimization, a single-metric model will be selected where $r$ is the area metric. If the design objectives are minimizing delay with a given area budget, a multi-metric model will be selected. Note that the \textit{number of classes} ($n+1$) is a \underline{fixed input} of the proposed framework, and the definition (QoR range) of each class is decided using a general model. For example, to define seven classes ($n$=6) in a single-metric model, it requires six determinators, \{$x_0$, $x_1$, ..., $x_n$\}. We define the six determinators using the \{5\%,15\%,40\%,65\%,90\%,95\%\} QoR results of collected labeled synthesis flows. For example, assuming 1000 labeled flows collected, $x_0$ is the $50^{th}$ least value of the select metric and $x_6$ is the $50^{th}$ largest value. Since the training dataset is updating incrementally, the definitions of classes may change dynamically. \textit{Angel-flows} and \textit{devil-flows} are the subset of the flows corresponding to classes \textit{\bf 0} and \textit{\bf n}. 

\begin{table}[!htb]
\centering
\scriptsize
\caption{Labeling the training flows based on synthesis QoR. $r$ and $r_i$ are the QoR metrics such as delay, area, power, etc.}
\label{tbl:labeling}
\begin{tabular}{|c|l|c|}
\hline
Single-metric      & Multi-metric & Class/Label    \\ \hline
$r \leq x_0$              &  $r_0 \leq x_0$, $r_1 \leq y_0$            & \textit{0}   \\ \hline
$x_{0}<r \leq x_1$ &  $x_0<$$r_0 \leq x_1$, $y_0<$$r_1 \leq y_1$           & \textit{1}   \\ \hline
$x_{1}<r \leq x_2$ &   $x_1<$$r_0 \leq x_2$, $y_1<$$r_2 \leq y_2$           & \textit{2}   \\ \hline
...                               & {\it ...}             & \textit{...} \\ \hline
$r > x_n$                 &    $r_0 > x_n$,$r_1 > y_n$           & \textit{n}   \\ \hline
\end{tabular}
\end{table}

\textcircled{2} {\bf Design and train CNN Classifier}. The second component is training a CNN classifier that predicts the classes of unlabeled flows. To train a CNN classifier, the training data, i.e., labeled synthesis flows, need to be represented in the matrix. We present a \textit{one-hot} modeling that represents synthesis flow in \textit{binary matrix}. This model and the CNN architecture are introduced in Section \ref{sec:onehot}.

\textbf{\textcircled{3} Output Angel-flows and Devil-flows}. The trained classifier will be used to predict the classes of a large number of \textit{un-tested} sample flows. Although we are only interested in the flows in classes \textit{\bf 0} and \textit{\bf n}, the classifier may label many flows in these two classes. However, for the synthesis perspective, selecting a small set of flows is sufficient. In this work, the \textit{angel-flows} and \textit{devil-flows} are selected from the flows labeled with \textit{\bf 0} and \textit{\bf n} with highest \textit{prediction confidence}. The details are included in Section \ref{sec:angel-devil}.


\subsection{CNN Classifier}\label{sec:onehot}

\subsubsection{One-hot Representation of Synthesis Flow}

In this section, the one-hot representation model of synthesis flow is introduced for $m$-repetition flow. The $non$-repetition flow can be represented using the same model. 

Let $\mathbb{M}$ be the binary matrix of a $m$-repetition flow $F$ with $\mathbb{S}$=\{$p_0$, $p_1$,..., $p_n$\} (see Definition 2). The number of transformations in $F$ equals to the length of the flow $\mathbb{L}$=$n \times m$ (see Remark 2). Let the $j^{th}$ synthesis transformation in $F$ be $p_i$, $j \leq \mathbb{L}$, $i \leq n$. Its $n$-by-1 binary vector representation is $\mathbb{V}_{j}$, where $i^{th}$ element is 1 and the other elements are 0. $\mathbb{M}$ is an $\mathbb{L}$-by-$n$ matrix such that its $j^{th}$ row is $\mathbb{V}_{j}$. 

\textbf{Example 3:} We illustrate the one-hot representation model using flow $F_0$ shown in Example 2, such that $\mathbb{S}$=\{$p_0$, $p_1$\} and $F$=$p_0 \rightarrow p_0 \rightarrow p_1 \rightarrow p_1$, $\mathbb{M}$ is an 4-by-2 matrix.
\vspace{-1mm}
\[
\scriptsize
\begin{bmatrix}
   1 & 0\\
   1 & 0\\
   0 & 1\\
   0 & 1
\end{bmatrix}
\]

\subsubsection{CNN Architecture and Training}

\begin{figure}[!htb]
\centering
\includegraphics[width=.4\textwidth]{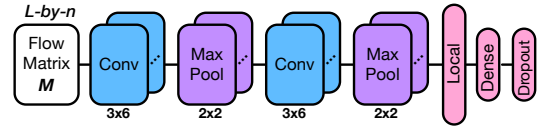}
\caption{CNN architecture used for synthesis flows classification.}
\label{fig:cnn_arch}
\vspace{-1mm}
\end{figure}

The input of the CNN are $\mathbb{L}$-by-$n$ binary matrices representing the synthesis flows. The CNN includes convolution, pooling, locally connected, dense and dropout layers. The \underline{kernel size} of the convolutional and pooling layers are shown in Figure \ref{fig:cnn_arch}. The dropout rate in the dropout layer is 0.4 to prevent the overfitting problem \cite{srivastava2014dropout}. Since our inputs are in one-hot representation, the loss function is computed using \textit{sparse softmax cross entropy} function. The output of the network comes from \textit{softmax} function. The number of kernels (filters) of convolutional layers are 200. The stride size of the convolutional and pooling layers are $1 \times 1$. 

Regarding the CNN architecture, two parameters have significant impacts on the prediction performance (accuracy): \textbf{a)} kernel size of convolutional layers and \textbf{b)} activation functions of convolutional and dense layers. Unlike most of the CNN classification applications, the $n$-by-$n$ kernel size does not perform well in classifying synthesis flows. We use $n \times 2n$ kernel size in this work. The reason is that there is only one non-zero element in each row of $\mathbb{M}$. Using $n \times 2n$ kernel could avoid computations over zero-matrix. The results of comparing the accuracy of the CNN classifier using 3$\times$6, 6$\times$6, and 6$\times$12 kernels are shown in Section \ref{sec:results} Figure \ref{fig:ksize_comp}.

The activation function of the nodes in the neural network defines the output of the nodes with a given set of inputs. In artificial neural networks, this function is also called the transfer function. The activation operations should provide different types of nonlinearities in the neural networks to solve Multiclass Classification problems. In general, there are two types of activation functions, including smooth nonlinear functions, such as \textit{Sigmoid, Tanh, Exponential Linear Units} (ELU) \cite{clevert2015fast}, \textit{Scaled Exponential Linear Units (SELU)}\cite{klambauer2017self}, etc., and smooth continuous functions, such as \textit{Rectified linear unit (ReLU)} \cite{nair2010rectified}, \textit{Concatenated Rectified Linear Units (CReLu)}\cite{shang2016understanding}, etc. We find that for classifying synthesis flows, the activation functions with nonlinearities perform better, such SELU and Tanh. The activation functions including ReLU, ReLU6, ELU, SELU, Softplus, Softsign, Sigmoid and Tanh, have been compared in Section \ref{sec:results} Figure \ref{fig:act_func_comp}.

Regarding the training process, the CNN classifier is trained specifically for each design as described in Section \ref{sec:overview}. Since the training data are collected incrementally, the CNN will be re-trained after every 500 new data points collected. The Mini-Batch \cite{orr2003neural} training strategy is applied in this work with batch size 5, i.e., simultaneously evaluated five training examples in each iteration. In this work, we have evaluated five different gradient descent algorithms, including Stochastic gradient descent (SGD), Momentum \cite{qian1999momentum}, AdaGrad \cite{duchi2011adaptive}, RMSProp \cite{tieleman2012lecture}, and Follow the regularised leader (FTRL) \cite{mcmahan2013ad}. The comparison result is included in Section \ref{sec:results} Figures \ref{fig:optimizer_area} and \ref{fig:optimizer_delay}. 

\subsection{Angel-Flows and Devil-Flows}\label{sec:angel-devil}

In this work, the outputs of the proposed framework are 200 \textit{angel}-flows and 200 \textit{devil}-flows. There are two steps for generating these flows. First, it uses the trained CNN classifier to predict the class of a large number of random flows. According to the classification rule (Table \ref{tbl:labeling}), the \textit{angel}-flows and \textit{devil}-flows will be selected from the $0$-class flows and $n$-class flows. The predicted class of a random flow is the class corresponding to the highest probability in the result of the CNN classifier coming from \textit{softmax} function. For example, assuming the output of the classifier (\# classes = 7) is \{$p_{0}=0.47, p_{1}=0.13, p_{2}=0.22, p_{3}=0.02, p_{4}=0.03, p_{5}=0.12, p_{6}=0.01$\}, where $p_i$ is the probability of a flow being class-$i$, then the predicted class is class-$0$. To minimize the errors in selecting the \textit{angel}(\textit{devil})-flows, our framework selects the flows with highest $p(0)$($p(n)$) \underline{within} the class-$0$(class-$n$) flows. 

\textbf{Example 4:} Let the prediction results in Table \ref{tbl:predict-example} be the prediction outputs of the CNN classifier of four synthesis flows. If two \textit{angel}-flows are required, $F_0$ and $F_1$ are selected and $F_4$ is eliminated. 

\begin{table}[t]
\scriptsize
\centering
\caption{Example of finalizing the \textit{angel}-flows.}
\label{tbl:predict-example}
\begin{tabular}{|l|l|l|l|l|l|l|l|}
\hline
 Flow  & $p_0$ & $p_1$ & $p_2$ & $p_3$ & $p_4$ & $p_5$ & $p_6$ \\ \hline
$F_0$ & \textbf{0.47} & 0.13 & 0.22 & 0.02 & 0.03 & 0.12 & 0.01 \\ \hline
$F_1$ & \textbf{0.51} & 0.12 & 0.01 & 0.09 & 0.17 & 0.08 & 0.02  \\ \hline
$F_2$ &   0.02 & \textbf{0.45} & 0.14 & 0.12 & 0.11 & 0.10 & 0.06   \\ \hline
$F_3$ &  0.12 & 0.03 & 0.17 & \textbf{0.62} & 0.01 & 0.02 & 0.03   \\ \hline
$F_4$ &  \textbf{0.35} & 0.23 & 0.09 & 0.02 & 0.13 & 0.17 & 0.01   \\ \hline
\end{tabular}
\vspace{-3mm}
\end{table}


%% file: results.tex
\begin{figure*}[!htb]
\centering
\begin{minipage}{0.26\textwidth}
  \centering
\includegraphics[width=1\textwidth]{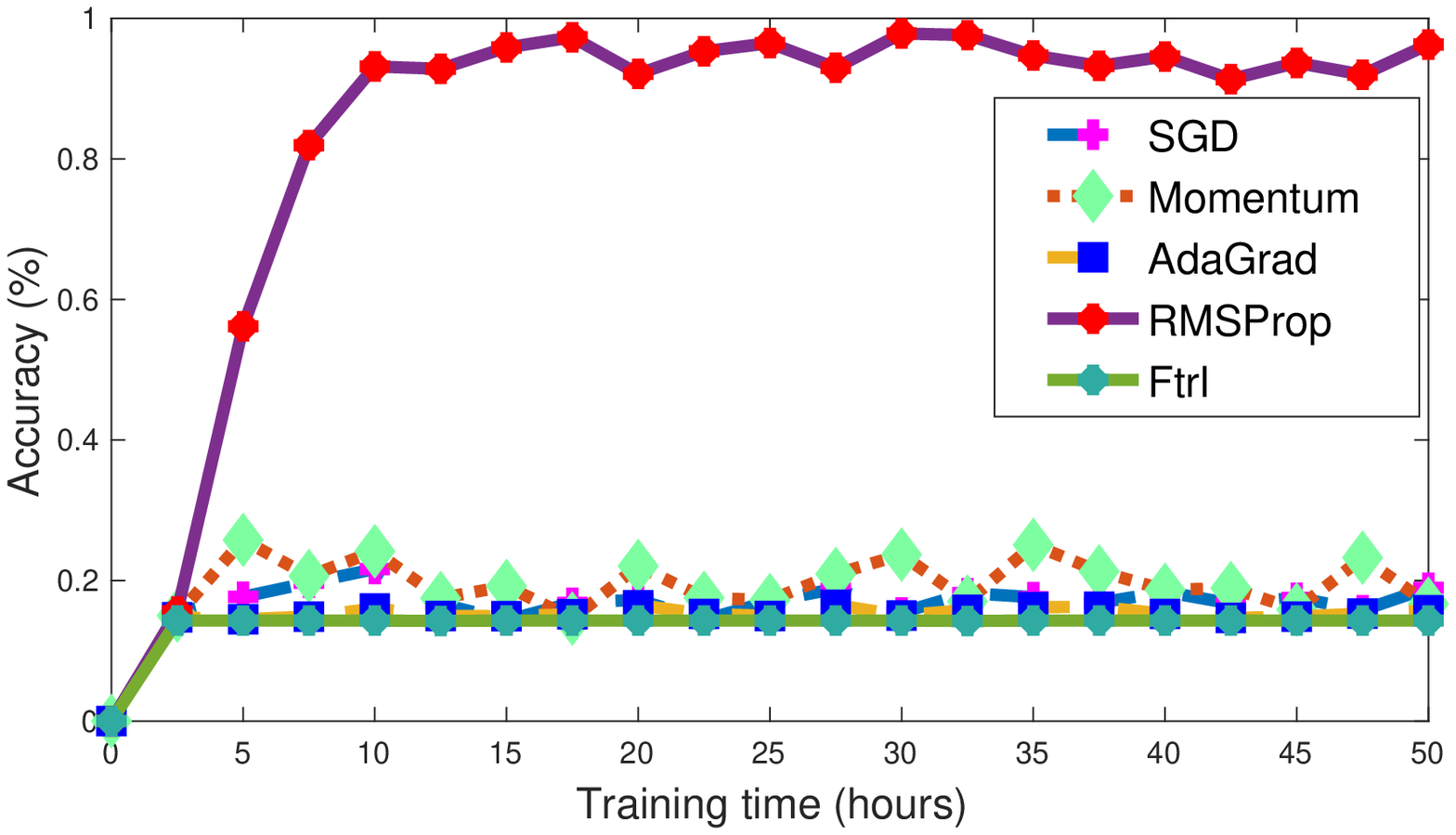}
\subcaption{Montgomery Multiplier}\label{fig:1a}
\end{minipage}%
\begin{minipage}{0.26\textwidth}
  \centering
\includegraphics[width=1\textwidth]{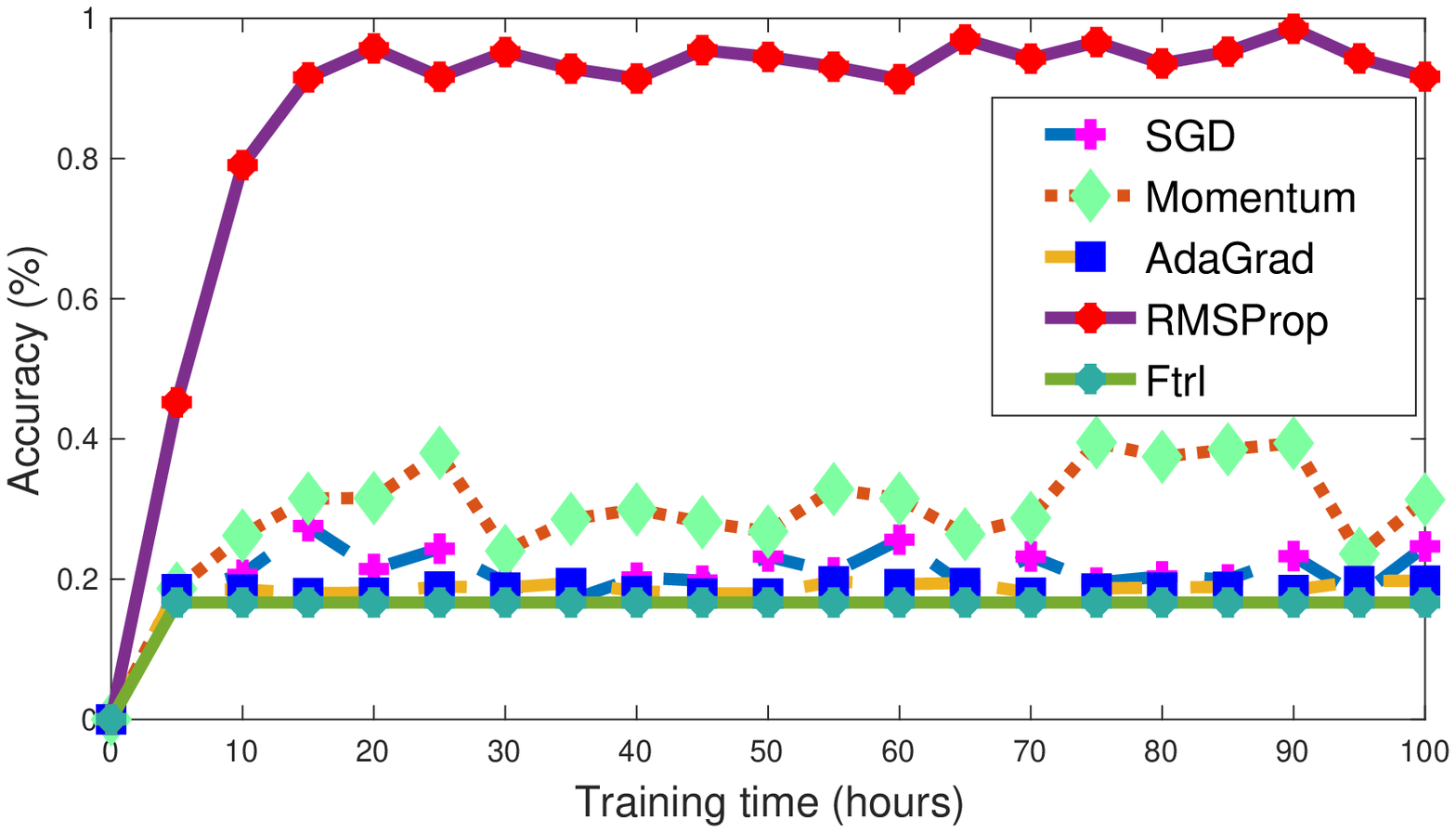}
\subcaption{AES Core}\label{fig:1b}
\end{minipage}%
\begin{minipage}{0.26\textwidth}
  \centering
\includegraphics[width=1\textwidth]{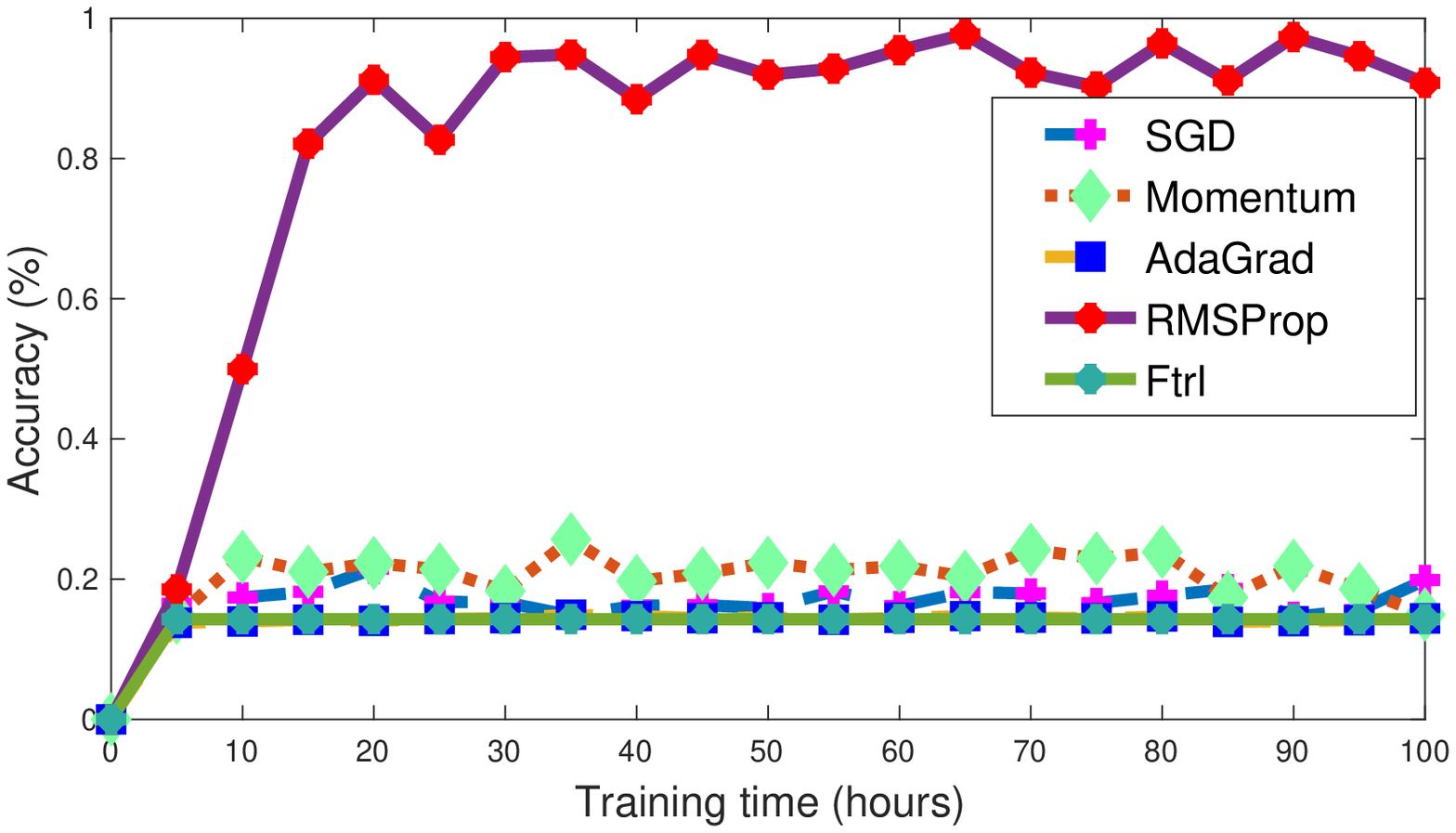}
\subcaption{ALU}\label{fig:1b}
\end{minipage}%
\caption{Evaluation of different gradient descent algorithms for generating the area-driven \textit{angel}/\textit{devil} flows.}
\vspace{-4mm}
\label{fig:optimizer_area}
\end{figure*}

\begin{figure*}[!htb]
\centering
\begin{minipage}{0.26\textwidth}
  \centering
\includegraphics[width=1\textwidth]{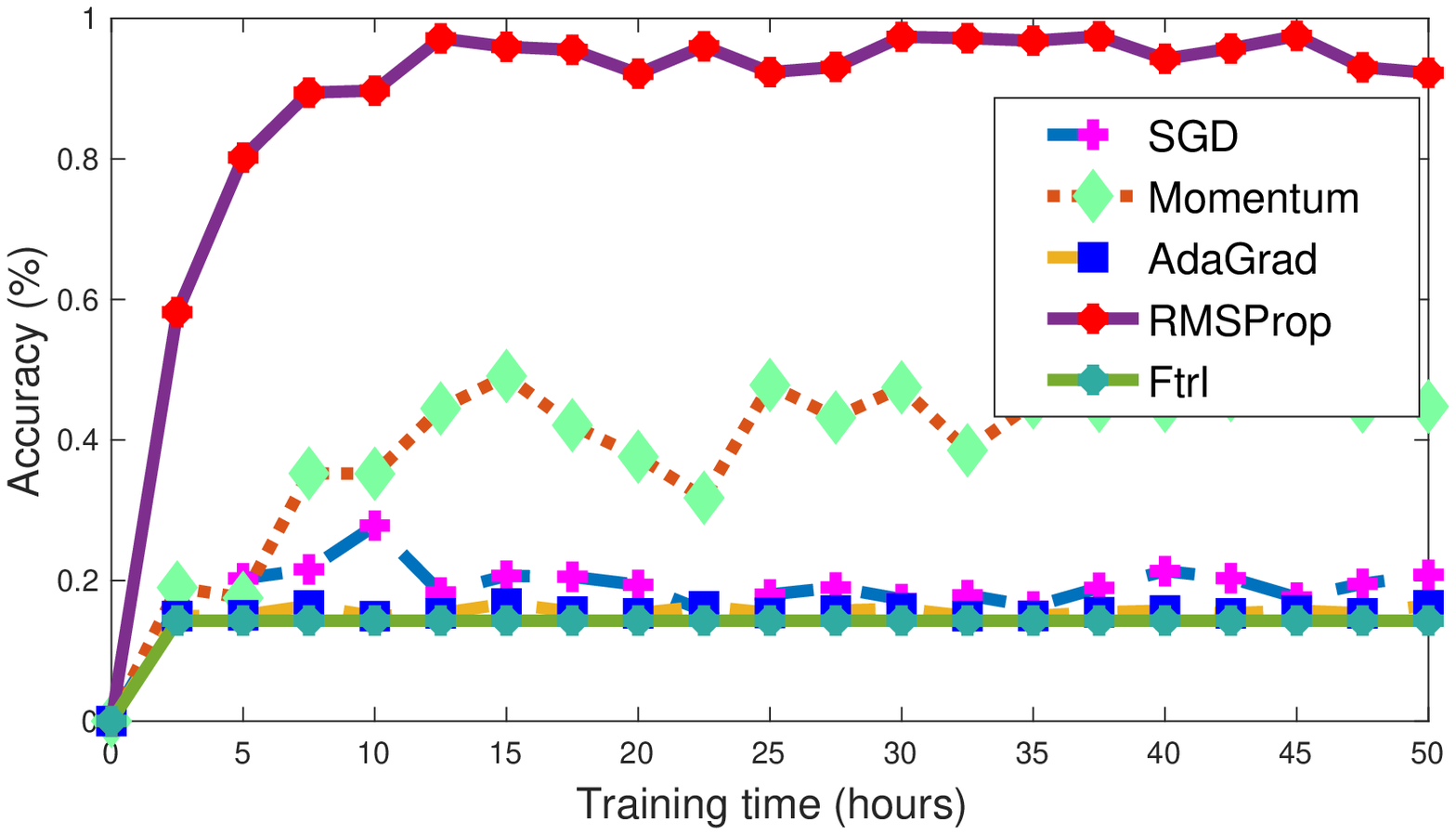}
\subcaption{Montgomery Multiplier}\label{fig:1a}
\end{minipage}%
\begin{minipage}{0.26\textwidth}
  \centering
\includegraphics[width=1\textwidth]{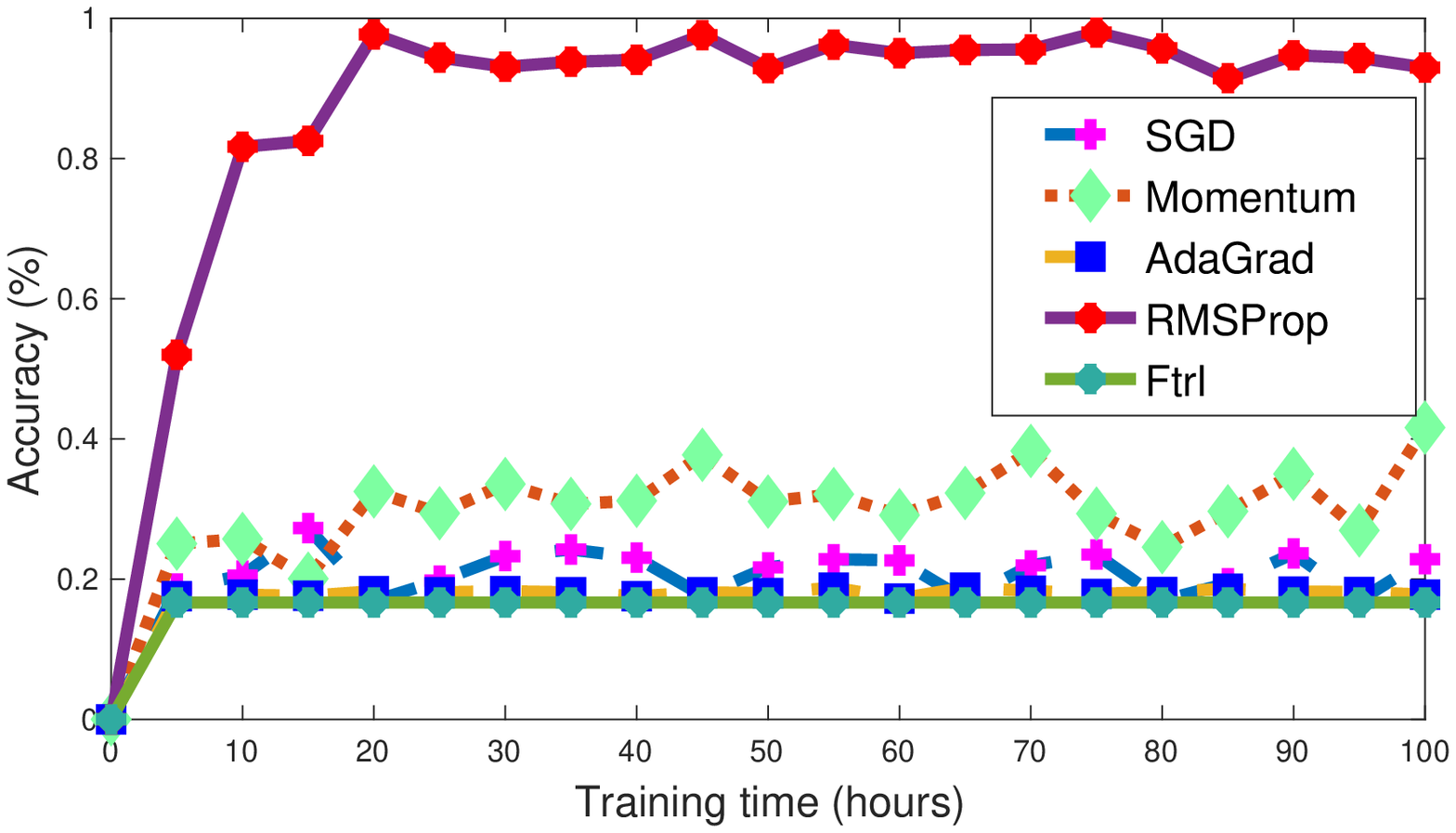}
\subcaption{AES Core}\label{fig:1b}
\end{minipage}%
\begin{minipage}{0.26\textwidth}
  \centering
\includegraphics[width=1\textwidth]{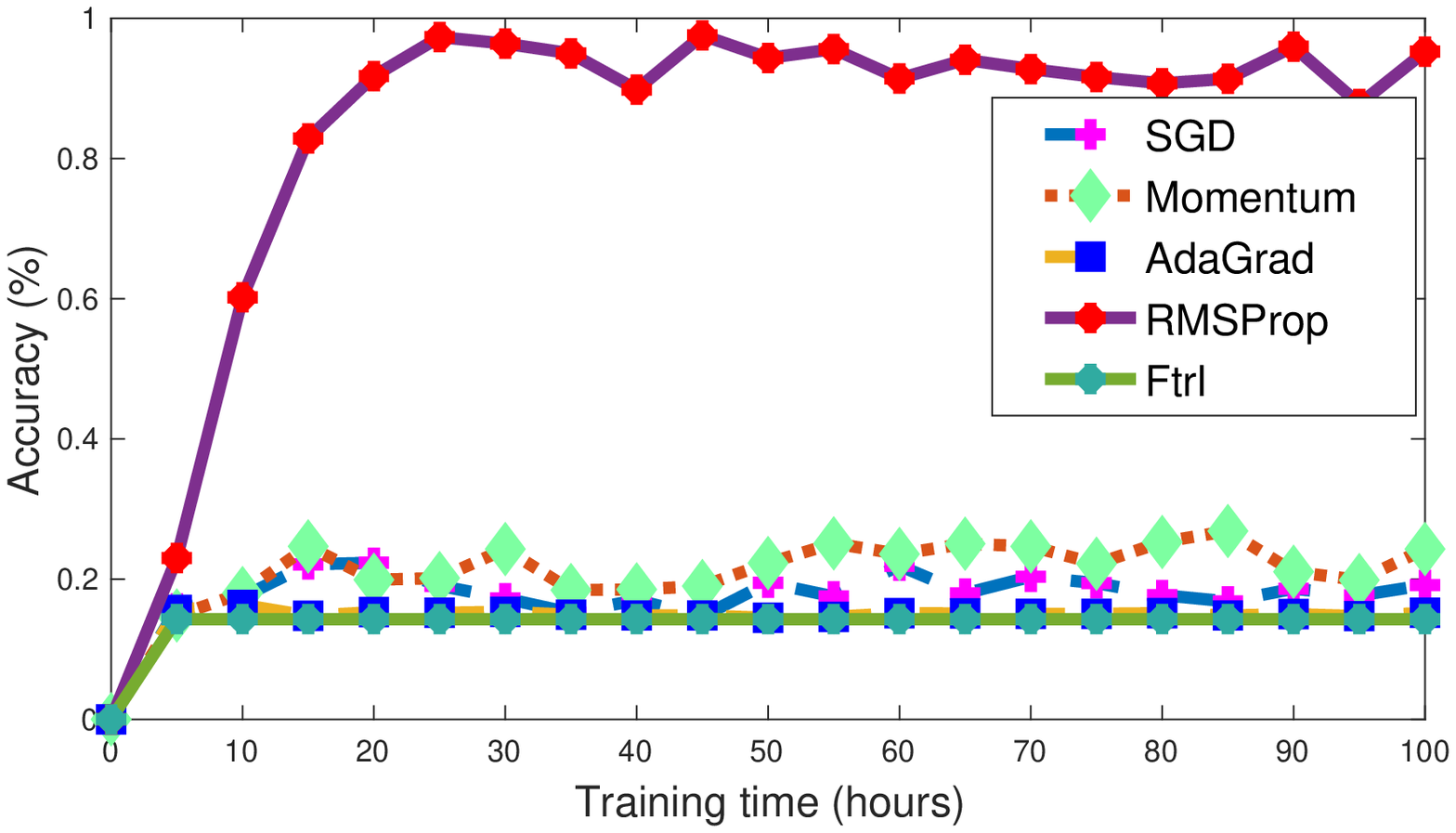}
\subcaption{ALU}\label{fig:1b}
\end{minipage}%
\caption{Evaluation of different gradient descent algorithms for generating the delay-driven \textit{angel}/\textit{devil} flows.}
\vspace{-4mm}
\label{fig:optimizer_delay}
\end{figure*}

\section{Experimental Results}\label{sec:results}

We demonstrate the proposed framework by designing logic synthesis flows Open-source synthesis framework ABC \cite{mishchenko2010abc}. Our framework is implemented in C++. The CNN classifier is implemented using Tensorflow r1.3 \cite{abadi2016tensorflow} using its C++API. The demonstration is made with three designs, including 64-bit Montgomery multiplier, 128-bit AES core \cite{opencore-web}, and 64-bit ALU \cite{opencore-web}. The goal is to generate 200 \textit{angel}-flows and 200 \textit{devil}-flows for area or delay optimization. We use the same setups shown in the motivating example (Section \ref{sec:motivating-example}). Thus, the synthesis flows will be $4$-repetition flows with six ABC synthesis transformations, $\mathbb{S}$=\{balance, restructure, rewrite, refactor, rewrite -z, refactor -z\}. The inputs of CNN classifier are $24$-by-$6$ matrices representing the synthesis flows using the one-hot modeling. These matrices are re-shaped to $12$-by-$12$ matrices for using two convolutional layers.

For generating the area- or delay-driven flows, we use the single-metric classification model (Table \ref{tbl:labeling}) where $r$ is the area/delay of the design. The number of classes is \underline{seven}. The six determinators are defined using \{ 5\%, 15\%, 40\%, 65\%, 90\%, 95\% \} of the area/delay results of the training flows.  The area and delay results are obtained after technology mapping with 14nm standard-cell library. The number of training flows is 10,000 and the number of sample flows for generating the final flows is 100,000. The experimental results are obtained using a machine with Intel Xeon 2x12cores@2.5 GHz, 256GB RAM, 2x240GB SSD and 2 Nvidia Titan X GPUs. 

The result section includes two parts. The first part contains the experimental results of training the CNN classifier. It consists of the evaluations of different gradient descent algorithms, various of convolutional kernel sizes and activation functions. Based on these results, we find the best settings for the CNN architecture and training strategy. Using these setting, we generate and evaluate the quality of generated \textit{angel}-flows and the \textit{devil}-flows. To evaluate the accuracy of the CNN classifier and the generated flows, we have explicitly collected the area and delay result by applying the 100,000 flows to the three designs. Hence, the true classes of the 100,000 sample flows are available for evaluation.

\vspace{-1mm}
\subsection{Results of Training CNN Classifier}

\vspace{-2mm}
\subsubsection{Gradient Descent Algorithms}

The results of training the CNN classifier using different gradient descent algorithms are shown in Figures \ref{fig:optimizer_area} and \ref{fig:optimizer_delay}. Figure \ref{fig:optimizer_area} includes the results of training for generating area-driven flows using five different algorithms, including Stochastic gradient descent (SGD), Momentum \cite{qian1999momentum}, AdaGrad \cite{duchi2011adaptive}, RMSProp \cite{tieleman2012lecture}, and Follow the regularised leader (FTRL) \cite{mcmahan2013ad}; Figure \ref{fig:optimizer_delay} includes results of generating delay-driven flows. The learning rate $\eta$=$0.0001$ and number of training steps is 100,000. The kernel size of convolutional layers is 6-by-12. In Figures \ref{fig:optimizer_area} and \ref{fig:optimizer_delay}, the $y$-axis represents the accuracy of prediction. Let $N_{angel}$ be the number of generated \textit{angel}-flows that their true class is class-$0$; let $N_{devil}$ be the number of generated \textit{angel}-flows that their true class is class-$6$. The accuracy is defined as following:
\[
accuracy = \{ (N_{angel} + N_{devil}) / 2 \times 200 \} / 2
\]
The $x$-axis represents the training time of our framework. Note that the training process of the 64-bit Montgomery multiplier is 2$\times$ faster than the other two designs. The reasons is that collecting the training dataset takes most of the runtime. The runtime of applying one synthesis flow to Montgomery multiplier is about 2$\times$ faster than the other two. The actually runtime for training the CNN classifier is about 3 - 5\% of the entire training time. As shown in Figures \ref{fig:optimizer_area} and \ref{fig:optimizer_delay}, the RMSProp \cite{tieleman2012lecture} outperforms other algorithms in classifying synthesis flows. The accuracy of the classifier in these six experiments reaches 95\% after 24 hours.

\subsubsection{Choice of Convolutional Kernel Size}

\begin{figure}[!htb]
\centering
\includegraphics[width=.37\textwidth]{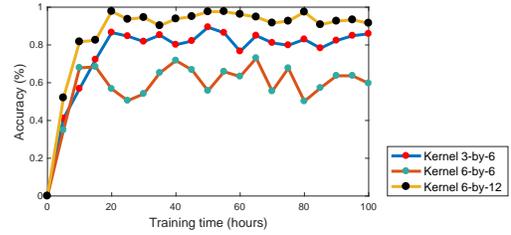}
\caption{Evaluation of three convolutional kernels. Test case: generating delay-driven flows for the 128-bit AES core.}
\vspace{-4mm}
\label{fig:ksize_comp}
\end{figure}

As mentioned in Section \ref{sec:onehot}, the size of the convolutional layer kernel has significant impacts on the CNN classifier. In Figure \ref{fig:ksize_comp}, three kernel sizes, 3$\times$6, 6$\times$12, have been tested using RMSProp algorithm \cite{tieleman2012lecture}, where the learning rate $\eta$=$0.0001$ and number of training steps is 100,000. The number of kernels at each convolutional layer is 200. The input design is the 128-bit AES core, and the objective is generating delay-driven flows. We can see that the kernel with size $n \times 2n$ (3$\times$6, 6$\times$12) perform much better than the $n \times n$ kernel (6$\times$6).


\subsubsection{Evaluation of Activation Functions}

\begin{figure}[!htb]
\centering
\includegraphics[width=.27\textwidth]{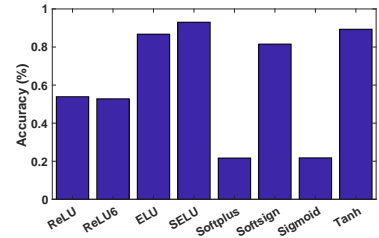}
\caption{Evaluation of different activation functions. Test case: generating delay-driven flows for the 128-bit AES core.}
\vspace{-5mm}
\label{fig:act_func_comp}
\end{figure}

For evaluating the performance of classifying synthesis flows using different activation functions, we set the learning rate $\eta$=0.0001, learning steps=100,000, convolutional kernel size is 6$\times$12, and use RMSProp to minimize the loss function. Figure \ref{fig:act_func_comp} includes the comparison of eight different activation functions, including ReLU, ReLU6, ELU\cite{clevert2015fast}, SELU\cite{klambauer2017self}, Softplus, Softsign, Sigmoid and Tanh. We can see that the ELU, SELU, Softsign and Tanh functions outperform the others, and SELU offers the best accuracy for generating delay-driven flows for the 128-bit AES core. Note that the accuracy of different activation functions varies on different datasets. In this work, SELU provides most reliable performance. 

\subsection{Quality of Generated Flows}

\begin{figure}[!htb]
\centering
\begin{minipage}{0.41\textwidth}
  \centering
\includegraphics[width=1\textwidth]{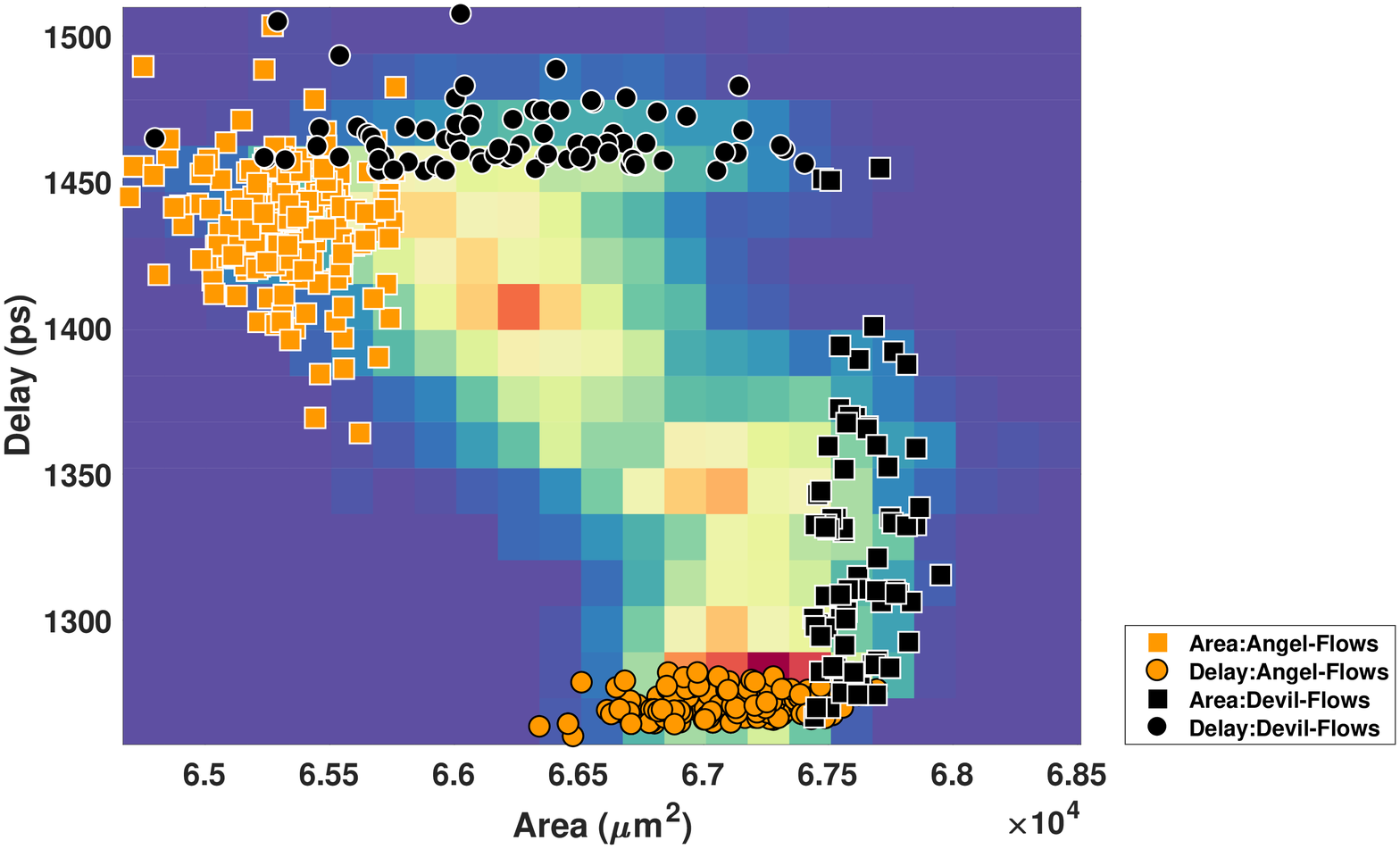}
\subcaption{Flows generated for 64-bit Montgomery multiplier.}\label{fig:1a}
\end{minipage}%
\\
\begin{minipage}{0.41\textwidth}
  \centering
\includegraphics[width=1\textwidth]{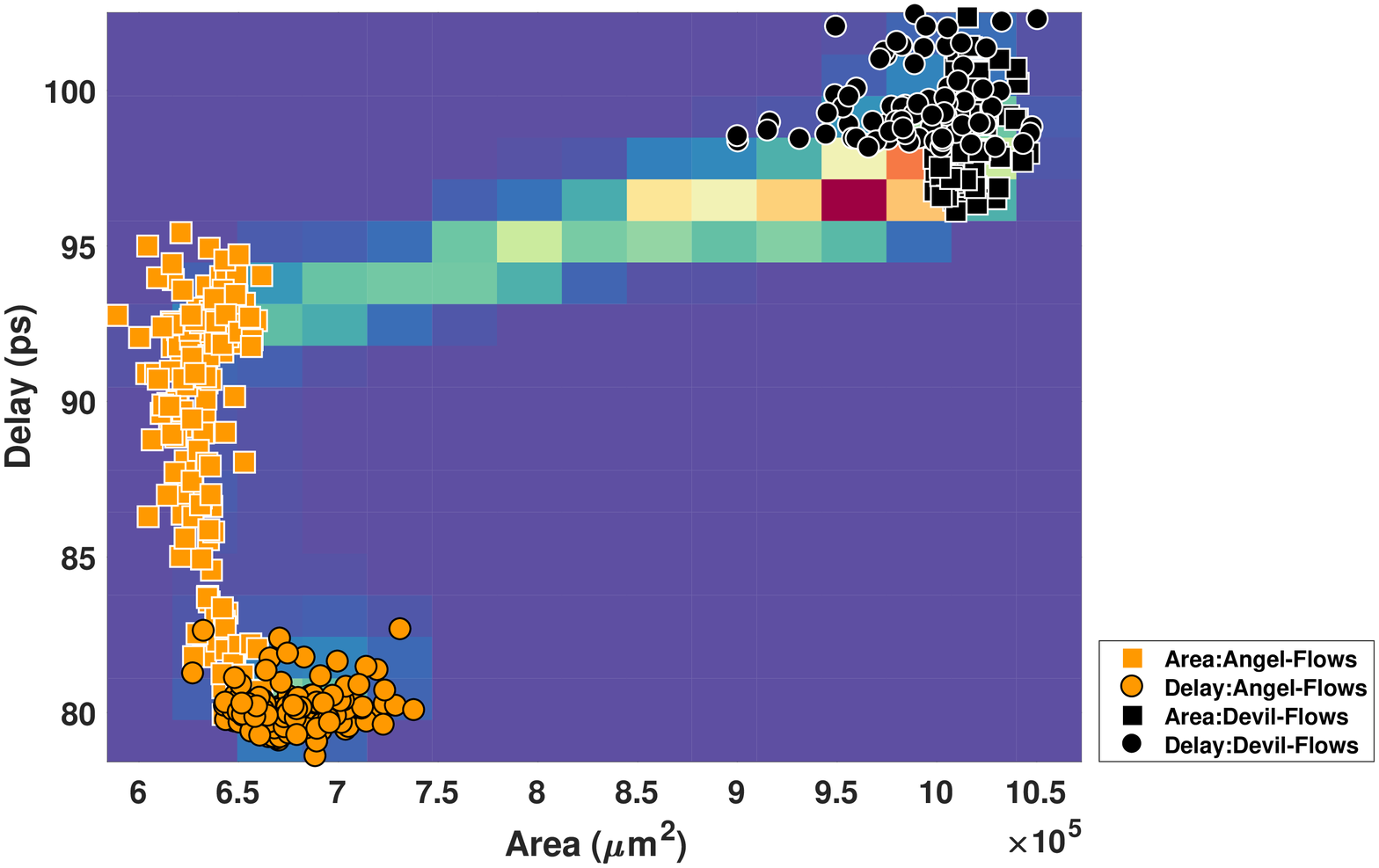}
\subcaption{Flows generated for 128-bit AES core.}\label{fig:1b}
\end{minipage}%
\\
\begin{minipage}{0.41\textwidth}
  \centering
\includegraphics[width=1\textwidth]{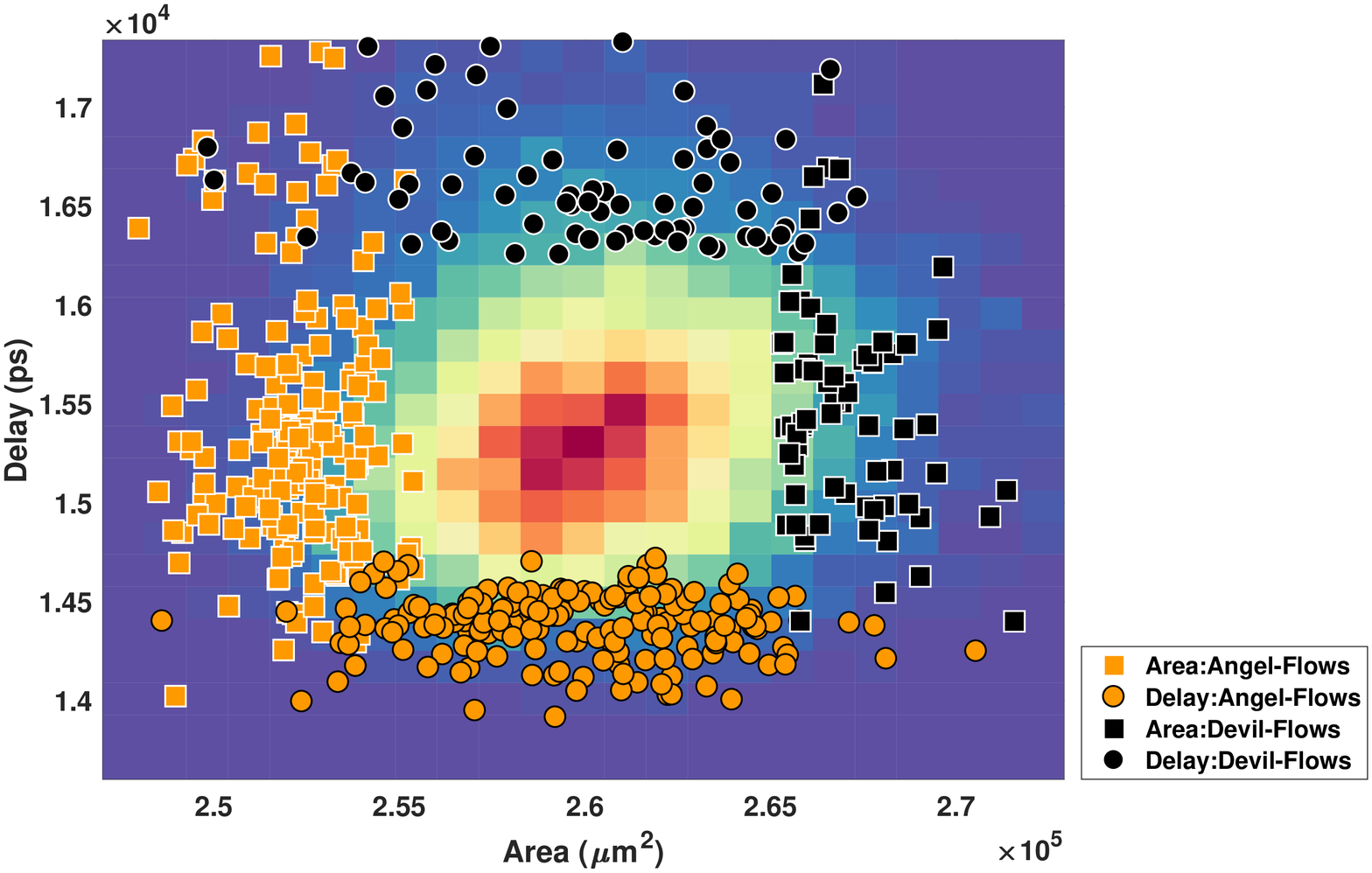}
\subcaption{Flows generated for 64-bit ALU.}\label{fig:1b}
\end{minipage}%
\caption{Quality of the generated ABC synthesis flows for 64-bit Montgomery multiplier, 128-bit AES and 64-bit ALU. }
\vspace{-4mm}
\label{fig:evaluation_flows}
\end{figure}

Finally, we evaluate the quality of the generated \textit{angel}-flows and \textit{devil}-flows. The results shown in Figure \ref{fig:evaluation_flows} are obtained using the following settings: number of training flows is 10,000; number of sample flows is 100,000; $\eta$=0.0001; learning steps is 100,000; activation function is SELU; gradient descent algorithm is RMSProp; convolutional kernel size is 6$\times$12. The four types of points shown in Figure \ref{fig:evaluation_flows} represent the area-delay result of \textit{area-angel-flows}, \textit{area-devil-flows}, \textit{delay-angel-flows} and \textit{delay-angel-flows}. The $y$-axis represents delay and $x$-axis represents area. The background of each sub-figures in Figure \ref{fig:evaluation_flows} is the 2-D distribution of the 100,000 sample flows\footnote{\small The 2-D distribution represents the distribution similarly to Section \ref{sec:motivating-example} Figure \ref{fig:motivation}, but with 100,000 data points.}. We can see that the generated area(delay) \textit{angel}-flows provide the best results in terms of area(delay), and the \textit{devil}-flows provide the worst results, among the 100,000 sample flows. For example, the data points of area-angel-flows of these three designs are clearly bounded with a certain area value. The total runtime for generating these flows takes 3-4 days. It is demonstrated that our framework can successively develop \textit{angel}-flows and \textit{devil}-flows.

\section{Conclusions}\label{sec:results}

This work presents a fully autonomous framework that artificially produces \textit{design-specific} synthesis flows without human guidance and baseline flows. We introduce a general approach for flow optimization problems by modeling into \textit{Multiclass Classification}. The one-hot modeling of iterative flows is proposed such that any flow can be represented using binary matrix. This approach is demonstrated by generating the best, and worst synthesis flows, using three large designs with 14nm technology. The future work will focus on artificially developing cross-layer synthesis flows to find the missing-correlations between logic and physical designs \cite{DBLP:conf/iccad/YuCSC17}.

\section{acknowledgement}
This project is funded by ERC-2014-AdG 669354 grant.